\documentclass[review]{elsarticle}
\usepackage{epsfig}
\usepackage{amssymb}
\usepackage{amsmath}
\usepackage[utf8]{inputenc}
\usepackage[T1]{fontenc}
\usepackage{graphicx}
\usepackage{caption}
\usepackage{subcaption}
\begin{document}
\begin{frontmatter}
\title{Diffusion of Innovation In Competitive Markets-A Study on the Global Smartphone Diffusion}

\author{Semra G\"und\"u\c{c}\footnote{e-mail:gunduc@ankara.edu.tr}}
\address{Ankara University Faculty of Engineering,\\
 Department of Computer Engineering G\"olba\c{s}{\i} Ankara Turkey}
        %\email{gunduc@ankara.edu.tr}
%\homepage{http://}
        
\begin{abstract}

In this work, the aim is to study the diffusion of innovation of two
competing products.  The main focus has been to understand the effects
of the competitive dynamic market on the diffusion of innovation. The global
smartphone operating system sales are chosen as an example. The
availability of the sales and the number of users data, as well as the
predictions for the future number of users, make the smartphone
diffusion a new laboratory to test the innovation of diffusion models
for the competitive markets.  In this work, the Bass model and its
extensions which incorporate the competition between the brands are
used.  The diffusion of smartphones can be considered on two levels:
The product level and the brand level. The diffusion of the smartphone
as a category is studied by using the Bass equation (Category-level
diffusion). The diffusion of each competing operating system (iOS and
Android) are considered as the competition of the brands, and it is
studied in the context of competitive market models (Product-level
diffusion). It is shown that the effects of personal interactions play
the dominant role in the diffusion process. Moreover, the volume of
near future sales can be predicted by introducing appropriate dynamic
market potential which helps to extrapolate the model results for the
future.

\end{abstract}


\end{frontmatter}

\section{\label{Introduction}Introduction}

%-------------------------

Diffusion of innovation studies has always been a significant part of
the marketing strategies when a new product is introduced to the
markets~\cite{Rogers:1962, Rogers:2003}.  The changes in technologies
and social behavior of the individuals affect the marketing strategies. Moreover, as the advanced technology products appear in the markets, more often similar products are introduced by different
brands. Competition between the similar products of different brands
can be both distractive or constructive. The existence of the competing
products increases the word of mouth which helps the growth of the
market potential. Hence, gaining a better understanding of the
diffusion of new products in competitive markets is growing interest
for both the scientists and the practicing economists. Predictions on
the conditions of success or failure of a new product in the
competitive markets~\cite{Rogers:2003, Mahajan:2000, Meade:2006} may
lead the changes of policies and even on the product specifications.

$\;$

%--------------------

The first mathematical model of the diffusion of innovation in the
literature is the Bass Model~\cite{Bass:1969}. The Bass model assumes
a single product diffusion in a homogeneous and fully connected social
network. The Bass model classifies the adoption characteristics of the
individuals into two different groups. The first group is the
individuals who decide to adopt the new product immediately after
being informed.  The external influences such as advertisement play
the role of the driving force. This group of individuals is named as
the innovators. The second group of individuals expects to see the
advantages of the product before adopting.  Hence in the adoption of
the second group of individuals personal interactions with already
adopted individuals or word-of-mouth is the crucial factor. The second
group is called imitators since their decision depends on the number
of adopted neighbors.

The Bass model contains three parameters. The probability of being
affected by the external influences is called the innovation parameter
$p$, while the number of imitators is controlled by the imitation
parameter $q$ . The potential market size is also another parameter
which is considered as a constant in the original form of the Bass
equation.

The Bass model well explains the diffusion of early technological
innovations, such as household goods, agricultural innovations,
automobiles, telephone, postal banking
systems~\cite{Hagerstrand:1968}. In the early days of technological
innovations, most of the time a brand could dominate a specific
market, and the diffusion process could be studied in a restricted
geographic area by the fixed size market potential assumption.

%-----------------------------

$\;$

%-------------------------------------

Recently, however, globalization, advancements in communication and
transportation technologies have changed the marketing considerations
completely.  Recent marketing observations indicate that almost always
similar new products are introduced as competing
brands~\cite{Savin:2005,Libai:2009,Krishnan:2000}. The rules of
diffusion are different when there exists competition between the
products.  In the competitive market case, more than one type of
internal influences exist.  The potential adopters are not only
influenced by the adopters of the product of interest, but also the
adopters of the competitive products take a role in their decision-making process. These new type influences are called cross-brand or cross-product effects~\cite{Hahn:1994,Givon:1995,Krishnan:2000,Laciana:2014}. 

Moreover,  the new products aim to reach the competitive global markets. Hence, the potential market size is dynamic~\cite{Mahajan:1978, Mahajan:1990}.  Even though there is no data-based evidence, Kim et al.~\cite{Kim:1999, Bulte:2004, Kauffman:2005} suggests that coexistence of competing products in the same market, indicate a potential for the products and help the growth of the markets. Hence in the competitive market case, the dynamics of the potential market size need to be considered together with the dynamics of the adopters.

%-------------------------------------

$\;$

%------------------

The existing literature on the studies of competitive markets varies
from investigations of the diffusion of new software under
piracy~\cite{Givon:1995} to diffusion of minivans~\cite{Krishnan:2000} or diffusion of mobile technologies and mobile
phones~\cite{Gruber:2001, Scharl:2005} to the diffusion of drugs of the
same active components~\cite{Guseo:2014}.  Modern communication systems and mobile phones are also attracted considerable interest. As an advanced version of mobile communication technologies, the diffusion of
smartphones and smartphone software are an exciting area of
diffusion of innovation studies.  The new smartphone operating systems, which
are iOS and Android have shown rapid success in the markets and
increased the sales rate and market size. The success of these new
operating systems has been the subject of various
studies~\cite{Tseng:2013, Wang:2016}. Tseng et al. ~\cite{Tseng:2013},
have used a model, in which scenario analysis, expert
opinion, Delphi and widespread diffusion of innovation models are used
in an integrated fashion to predict feature market shares of the smartphone operating systems. Wang et al., have also studied competition in
the smartphone markets~\cite{Wang:2016}. The Lotka-Volterra equation
which is initially introduced for the competition of species is used
for the short-term predictions for market shares of iOS and Android
operating systems~\cite{Wang:2016}. None of these studies have the power to
predict the expansion of the potential market size. The relation between
the time-dependent the potential market size and the market shares of
competing products is the relatively less known area of
research~\cite{Parker:1994, Peres:2010, Guseo:2012, Guseo:2014}.

%-----------------
$\;$
%-------------------

The aim of the present work is threefold: The First one is to
determine the values of external and internal influence parameters of
the smartphone operating systems. The parameter values of previous
technological innovations already obtained and listed. Hence a
comparison of the innovation and imitation parameters of the
smartphone operating system markets with the existing parameters will
be illuminating. The second aim is to establish a functional form for
the time dependence of the potential market size which has the power
to predict future sales. Finally, the third one is the calculation of
the growth of the potential market size due to the competition of the
brands. In modern societies, markets are dynamic; market growth is a
function of technological utility, economic and social considerations
as well as social prestige. Despite the difficulty of writing an
explicit form of the market potential regarding various social and
economic parameters parametric market potential can be used for the
predictions.  In this sense, the present work is a comparative study
of the effects of fixed and the different forms of dynamic market
potentials~\cite{Peres:2010,Guseo:2014}.

$\;$

In the computation of the diffusion of innovation of the smartphone operating systems, a two-stage process is adopted. As the first step, the category level diffusion parameters are obtained by using analytical and numerical solutions of the Bass equation for both time-independent and dynamic market situations. As the second step,
product-level internal, external and competition parameters of the diffusion of innovation are obtained by fitting the data to the analytical and numerical solutions of coupled differential equations-Extended Bass equations.  The parameters of the models obtained by fitting the available smartphone sales and user data~\cite{GlobalSales, WorldWideSmartphoneUsers} to the
solutions of the Bass and generalized Bass equations.

%------------------

$\;$

%--------------------<
The smartphone operating systems sales is a model for competitive
markets, and it is a representative of modern diffusion processes in
which global availability and fast information diffusion play an essential role~\cite{DeGusta:2012}.  For this reason, smartphone
diffusion has some properties which are Sui Generis. First of all, it is straightforward to isolate a group of products from others which does not serve the same purpose. The smartphone operating systems, regardless the brands,  constitutes the category.  Secondly, only the products of the two brands dominate the whole market. Hence the category has two competing products. Thirdly the diffusion of both, information and products are global and show very little difference in vast geographic distances. This property of the smartphone category
enables the use of the aggregate models: The category-level diffusion
of innovation can be studied by using the famous Bass
equation~\cite{Bass:1969}, while alternative models can be used to
study the competition between different products. In general, the
models which aim to explain the dynamics of product-level diffusion
are the generalizations of the Bass model.  Krishnan, Bass, and Kumar
~\cite{Krishnan:2000} introduced a model which connects category-level sales with product-level sales. The model introduced by
Guseo~\cite{Guseo:2009, Guseo:2015} has the same property. Muller at.,
al formulated the one which has fewer limitations compared with the
others ~\cite{Libai:2009}. Models which establish the direct
relation between the category-level and brand-level diffusion
processes also exist ~\cite{Hahn:1994, Givon:1995, Krishnan:2000, Guseo:2015}.

%--------------------<

$\;$

%----------------<
Many already mobile phone producing companies have started the
smartphones revolution. Starting from the first quarter of 2009,
quarterly global sales~\cite{GlobalSales} and global market
shares~\cite{GlobalShares} data for all smartphone operating systems
are available. Immediately after the introduction of smartphone idea
two new operating systems which are introduced in 2007 and 2008 started to dominate global markets.  Recently only these two operating
systems are dominating the global markets ( in 2017 99.6
\%)~\cite{AndroidAndiOS}. The number of actively used
devices~\cite{AndroidAndiOS,NumberOfActiveiOSUsers,NumberofActiveAndoidUsers},
announced by producers of both operating systems. Apart from the sales
and shares data, forecasts on the number of smartphone users until
2020 are also available~\cite{GlobalSales,WorldWideSmartphoneUsers}. Such information eases the modeling of data and enables to compare the
prediction capabilities of different models.

%----------------<

$\;$

%---------------<
In this work, the Bass equation has been used to calculate category
level diffusion of smartphone operating systems. The Bass equation
with constant market potential size is shown to capable of explaining
the existing sales data but predicts that the sales will saturate
after 2017. Different time-dependent potential market models eliminate
this shortcoming, and for the category level sales, the Bass model
predictions and predictions based on social and economic
considerations~\cite{WorldWideSmartphoneUsers} match well. Brand level
sales predictions require the use of generalized Bass model type
aggregate models. Three such models are considered by using both
constant and time-dependent potential market
size~\cite{Libai:2009, Krishnan:2000, Guseo:2015}. The existing data
provide an excellent platform to test various diffusion of innovation
models under competition. The results have shown that the smartphone
markets are still in their growth phase.

$\;$
The following section is devoted to the discussions on the existing competitive market models. In the third section, 
the smartphone operating system sales data is considered. In this section also the diffusion of innovation parameters are obtained by using the smartphone sales data, future predictions and a mathematical model of the competing markets.
 In the last section discussions of the
results and conclusions are presented.

%---------------<

\section{\label{Models}Innovation of Diffusion Models}

The Bass equation~\cite{Bass:1969} defines the diffusion of innovation
in terms of three parameters. The adoption rate at time $t$ is given by the
equation,

\begin{equation}
  \label{BassEquation}
\frac{dN(t)}{dt} = (p + q \frac{N(t)}{M(t)} ) (M(t) - N(t))
\end{equation}

%------------------------------VVVV--------------------
where, $p$, $q$ and $M$ are the external (innovation) and internal
(imitation) influence, and the market potential size parameters
respectively. $N(t)$ is the number of adopters at a given time $t$. In
the original Bass equation (Eqn.~\eqref{BassEquation}), the relation between the already adopted and the potential adopters, are
represented by a linear relation and the market potential  size
$M(t)=M$ is fixed. Hence market potential size is an upper bound of
the diffusion process.

$\;$ %---------------------AAAA---------------------
%------------------VVV------------------------
If the same market is shared by more than one product, the competition
changes the dynamics of the diffusion. Adopters of competing products
affect all potential adopters. The existence of competing products
requires the introduction of cross-product influences. The
influence of the adapters on the potential adopters of competitive
products may be positive or negative. Hence competition may even
increase the sales of competing products. The total sales of all
brands constitute the category; thus the market gain of one product does not directly mean a market loss of the other.

$\;$

The simplest model of competition is introduced by Krishnan, Bass, and Kumar~\cite{Krishnan:2000}. In this model category-level as well as the product-level diffusion equations are simple Bass equations. The
coupling between the simple Bass equations are satisfied with the restriction that the product level external, $p_i$, and internal,
$q_i$, parameters add up to the category level external, $p_c$, and
internal, $q_c$, parameters.  Krishnan-Bass-Kumar model (KBKM) assumes
that the potential adopters, $M(t)-N(t)$, and market potential size
$M(t)$, are common for both category and product level equations.
%----------------------------AAAA-----------------------------
\begin{eqnarray}
  \label{ExtendedBass}
  \frac{dN(t)}{dt} &=& \left (p_c + q_c \frac{N(t)}{M(t)}\right) (M(t) - N(t)) \nonumber\\
  & & \\
    \frac{dN_i(t)}{dt} &=& \left (p_i + q_i \frac{N(t)}{M(t)} \right) (M(t) - N(t)) \;\;\;{\rm where}\;\;\; i=1 \dots k \nonumber
\end{eqnarray}

Here, $k$ is the number of products which are sharing the same market
and $p_c$, $p_i$ and $q_c$, $q_i$ are the category-level and product
level external and internal influence parameters respectively.

$\;$

The constraint relations,
\begin{eqnarray}
  \label{Restrictions}
  p_c &=& \sum_i^k p_i \nonumber \\
  q_c &=& \sum_i^k q_i  \\
  N(t)&=& \sum_i^k N_i(t)\nonumber
\end{eqnarray}
%-------------------VVV------------------------
ensures the correct inter- and cross-product level influences. Here,
$N(t)$ is the total number of adopters which is the sum of the
adopters of all competing products, $N_i(t)$, at the market.

$\;$
The main disadvantage of the Bass-Kumar model~\cite{Krishnan:2000} is
that the relative effects of the influences on the adopters of the
competing products are not explicitly expressed in the model.

$\;$

Guseo and Guidolin~\cite{Guseo:2009} proposed a model,  for the market
potential growth. In the same work, the Bass model is extended by an
additional correction term to account a self-reinforcing effect of
the dynamic markets (Guseo and Guidolin Model, GGM).
%---------------------------AAAA---------------------
\begin{equation}
\label{Guseo1}
\frac{dN(t)}{dt} = \left( p + q \frac{N(t)}{M(t)} \right)
\left( M(t) - N(t) \right) + N(t) \frac{M^{\prime}(t)}{M(t)}
\;\;\; p, q > 0; \;\;  t \ge 0.
\end{equation}
%-------------------------VVVVVVVVV------------------------

where $p$ and $q$ are external and internal influences, $N(t)$ and
$M(t)$ are the number of adopters and dynamic market size at time $t$.
When the market potential is constant,  equation \eqref{Guseo1}
becomes Bass Equation.

$\;$

Later, Guseo and Mortarino have extended equation \eqref{Guseo1} to
describe the diffusion of two competing
products~\cite{Guseo:2014,Guseo:2015}.  Two equivalent forms of
coupled differential equations and their analytical solution is
presented in ref~\cite{Guseo:2015}. The model proposed by Guseo and
Mortarino will be named GMK hereafter.

%-------------------AAAA-------------------------

One of the equivalent models of competitive markets is given as,

\begin{eqnarray}
  \label{Guseo2}
  \frac{dN_1(t)}{dt} &=& \left (p_1 + q_1 \frac{N(t)}{M(t)}+ \delta
  \frac{N_1(t)}{M(t)}\right) \left(M(t) - N(t)) + N_1(t)
  \frac{M^{\prime}(t)}{M(t)} \right)\nonumber\\
  & & \\
  \frac{dN_2(t)}{dt} &=& \left (p_2 + (q_2-\delta) \frac{N(t)}{M(t)}+ \delta
  \frac{N_2(t)}{M(t)} \right) \left( M(t) - N(t)) + N_2(t) \frac{M^{\prime}(t)}{M(t)}\right) \nonumber
\end{eqnarray}

where the parameters $p_1,P_2,q_1,q_2 $ and $\delta$ controls the
diffusion and the competition between the products. One of the most
intriguing features of the equation \eqref{Guseo2} is that addition of
two equations gives the category level diffusion equation of the
products. Hence, GMM is also equivalent to KBKM.
$\;$

Interactions between the adopters and potential adopters of competing
products are also studied in alternative models. Majority of these
models are generalizations of the Bass model.  Libai formulates the
one which has fewer limitations compared with the others,

\begin{eqnarray}
  \label{CrossProductEffects}
  \frac{dN_i(t)}{dt} &=& (p_i +  \sum_j q_{i,j}  \frac{N_j(t)}{M} ) (M - N(t)) \;\;\; i,j = 1,\dots, k
\end{eqnarray}

here $k$ is the number of competing products, $p_i$, and $q_{i,j}$ are
the innovation and imitation parameters of the $i^{\rm th}$
product. The model introduced by Libai, Muller and
Peres~\cite{Libai:2009} hereafter will be called LMPM.  In this model
for each product, there exist more than one imitation
parameters. $q_{i,i}\;\;i,j=1,\dots,k$ and $q_{i,j},\;\;i,j=1,\dots,k
; j\ne i$ are the inter-product and cross-product imitation
parameters. The cross-product parameters, $q_{i,j}$, are the measure
of the degree to which adoption of the $i^{\rm th}$ product is
affected by the adopters of the $j^{\rm th}$ competing product
~\cite{Peres:2010}.

$\;$

The dynamic market growth is an interesting topic. After the
introduction of the product, new demand may be created and time-dependent
increase in the market size can be observed. Various form of dynamic
market growth is introduced in the literature.

The simplest market growth forms are the exponential time evolution of
the market size $M(t)$~\cite{Sharif:1981,Mayer:1999,Centrone:2007},

\begin{equation}
\label{MarketGrowth1}
M(t) = M(0)\exp(\delta t) \;\;\;{\rm where}\;\;\delta > 0.
\end{equation}

The Bass equation with exponentialy growing market potential is also
solved analitically~\cite{Sharif:1981}.

$\;$
An other form is introduced by Namwoon ~\cite{Namwoon:1999}:
\begin{equation}
  \label{MarketGrowthx}
M(t) = M(0)~\left(1-\exp(-\delta C( t))\right)
\end{equation}
where $M_0$ is the maximum potantial market size, $C(t)$ is the
number of competitors at time $t$.

In the smartphone operating system sales case, there are only two
competitors in the long term, the form used by Namwoon et al is
simplified to parametrize the time dependence of the market growth, in
this work.

\begin{equation}
  \label{MarketGrowth2}
M(t) = M(0)~\left(1-\exp(-\delta t)\right)\;\;\;\delta > 0
\end{equation}
Here, $\delta$ is the time constant of the market
growth.

Recently Guseo and Guidolin have introduced a new form of market
potential growth~\cite{Guseo:2009}. This form is obtained diffusion
model based on cellular automata dynamics. The introduced form of the
growth is basically the same as the Bass model : Growth of the awareness
of the potential adopters grows the market potential. The obtained
form,

\begin{equation}
   \label{MarketGrowth3}
  M(t) = M_0 \sqrt{ Z(t) }
  \end{equation}
where,
\begin{equation}
  Z(t) = \frac{1-\exp]\left[-(p_c+q_c)t \right]}{1 + q_c/p_c \exp]\left[-(p_c+q_c)t \right]}
\end{equation}
is used in studies of diffusion of competing drugs~\cite{Guseo:2015}.

$\;$

Equations \eqref{ExtendedBass} - \eqref{CrossProductEffects} are
solved under the assumptions that the potential market size is
constant and the potential market size is growing in time. The
consequences of having constant market potential size and time
dependent growth in the market potential size are very important,
which directly effects the prediction power of the models.

$\;$

In the following section the models will be applied to smartphone
operating system sales data.  The relative importance of diffusion
parameters and effects of the competition and the diffusion processes
will be discussed.

\section{\label{Results}Results and Discussions}

In this section the prediction power of different diffusion of
innovation models and the effect of market growth will be tested by
obtaining the parameter values from the data.  The available data are
in two groups: The first group consists of quarterly sales of
competing brands which are obtained from statista~\cite{GlobalSales},
the second set is the number of actively used devices starting from
2016 which also include predictions until
2020~\cite{WorldWideSmartphoneUsers}.

\begin{figure}[h]
    \centering
        \includegraphics[width=0.7\textwidth]{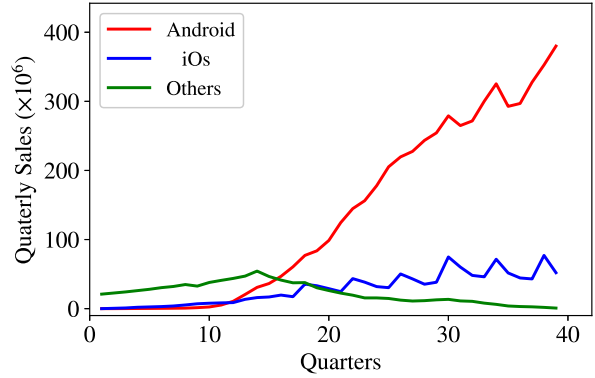}
    \caption{Global sales of smartphones operating systems starting
      from the first quarter of 2009}
    \label{fig:OperatingSystemsSales}
\end{figure}

Figure~\ref{fig:OperatingSystemsSales} shows that the global sales
increase at every quarter while the contributions of the brands other
than iOS and Android are rapidly decreasing.

\begin{figure}[h]
    \centering
     \begin{subfigure}[b]{0.47\textwidth}
        \includegraphics[width=\textwidth]{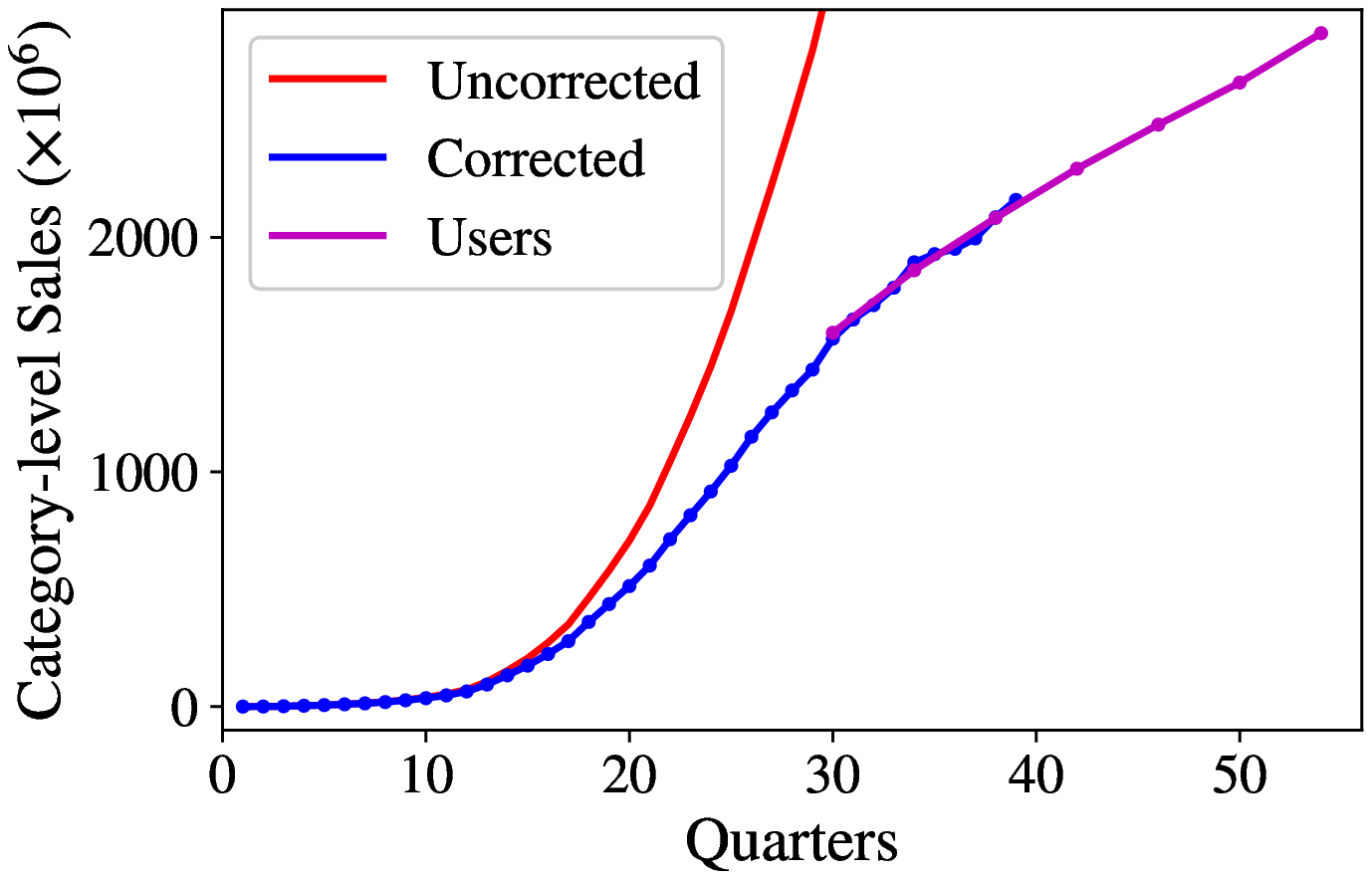}
        \caption{Cumulative category-level sales}
        \label{fig:SalesCategory}
    \end{subfigure}
    \begin{subfigure}[b]{0.47\textwidth}
        \includegraphics[width=\textwidth]{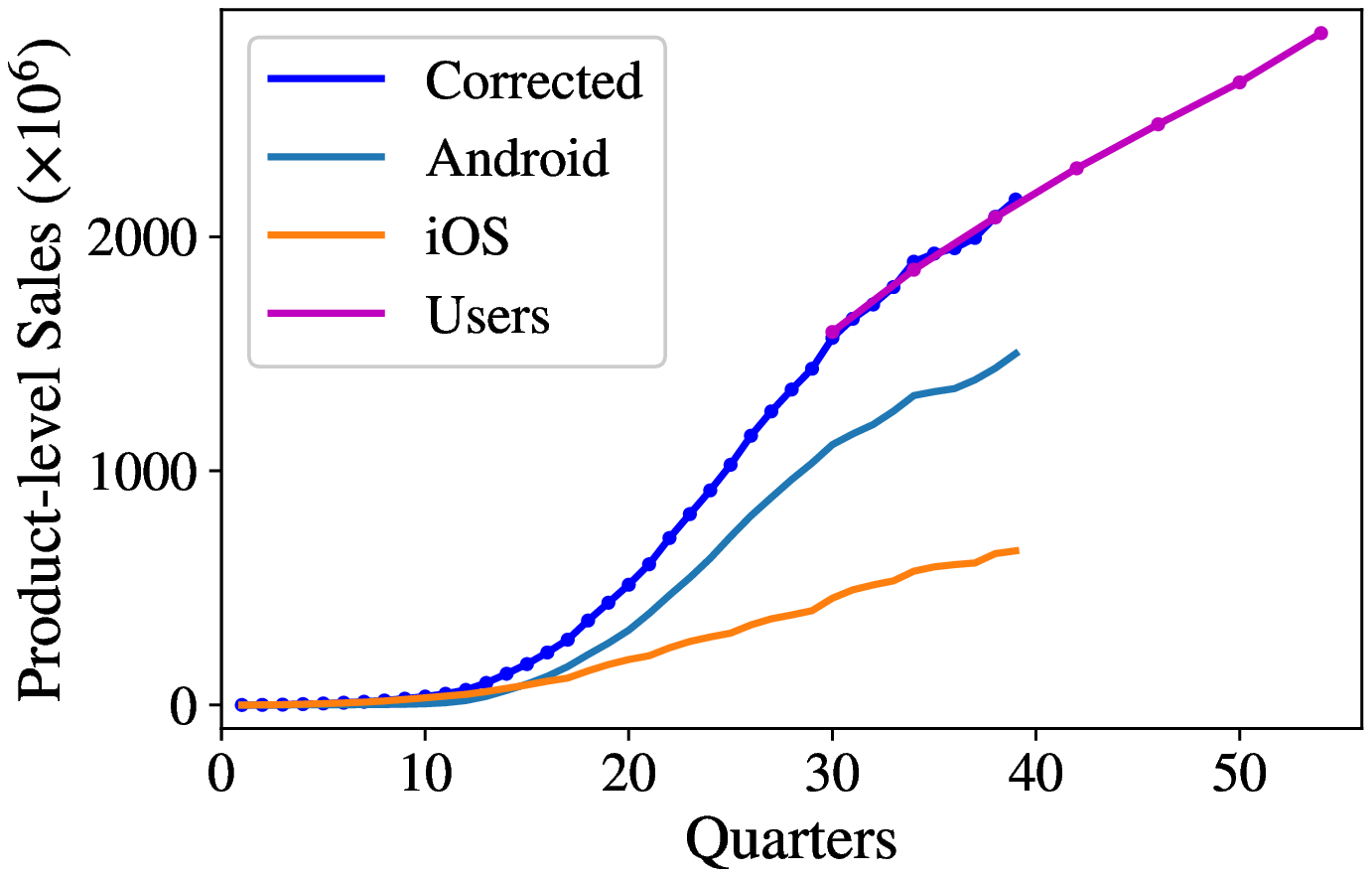}
        \caption{Cumulative product-level sales }
        \label{fig:SalesProduct}
    \end{subfigure}  
    \caption{Global cumulative sales of smartphones with iOS and
      Android operating systems starting from the first quarter of
      2009.}\label{fig:CumulativesalesOperatingSystems}
\end{figure}

The cumulative smartphone sales data which is necessary for comparison
with the model prediction can be obtained from quarterly sales as,

\begin{equation}
  \label{CumulativeSales}
S_{i}^a =\sum_{j=1}^i  s_j^a 
  \end{equation}

where $s_j^a$ and $S_i^a$ represent sales and cumulative sales of
brand $a$ at the $j^{\rm th}$ time slice respectively. The global
cumulative sales (Eqn.~\eqref{CumulativeSales}) obtained from the
data, figure~\ref{fig:OperatingSystemsSales}, are much higher than the
number of existing smartphones~\cite{WorldWideSmartphoneUsers}. This
discrepancy can be explained by considering relatively short active
lifespan of smartphones. To obtain sensible results, active product
lifespan must be taken into consideration. The difference can be seen
in Figure~\ref{fig:CumulativesalesOperatingSystems}.

The proposed empirical model assumes that every year a fraction of the
existing devices are becoming unusable. Hence this fraction must be
reduced from the cumulative sum

\begin{equation}
  \label{EmpricalCorrection}
  S_{i+1}^a = (1-f(i)) S_i^a + N_{i+1}^a
\end{equation}

where $f(i)$ is a time dependent factor representing the fraction of
the devices that are dropped from the sum. For simplicity the form of
the function is assumed linear,

\begin{equation}
\label{reductionterm}  
f(t) = f_0 + f_1 \times t.
\end{equation}

The parameters of the reduction function (Eqn.~\eqref{reductionterm})
are obtained by fitting the processed cumulative sales data and the
actual users
data~\cite{WorldWideSmartphoneUsers,AndroidAndiOS,NumberOfActiveiOSUsers}.
The best matching form (Equations~\eqref{EmpricalCorrection} and
\eqref{reductionterm} ) is obtained by an empirical fitting
process. The obtained coefficients $f_i^a$ are given by
table~\ref{parameters}.

\begin{table}[h!]
  \centering
  \begin{tabular}{|l|ll|}
    \hline
    &  $f_0^{(2)}$ & $f_1^{(2)}$ \\
        \hline
  Android &  $0.110$ & $0.003$ \\
  \hline
  iOS     & $0.025$ & $0.001$\\
  \hline
  \end{tabular}
  \caption{Parameters of empirical correction function.}\label{parameters}
\end{table}

The category-level cumulative numbers are calculated for each quarter
by adding product-level cumulative numbers. Figure
\ref{fig:SalesCategory} show the cumulative numbers without and with
the correction term. The number of active smartphone users are also
presented for comparison.

\subsection{Model Applications}

The diffusion of smartphones is studied at two levels: category and
brand levels. First, the category level diffusion is discussed. The
Bass equation~\eqref{BassEquation} and the model introduced by Guseo
and Guidolin~\cite{Guseo:2009} with different dynamic market
potentials are employed to study the internal and external effects on global smartphone diffusion. Second, three different but related
competitive market models with dynamic market potentials are tested on
the smartphone data. The global quarterly sales data indicate no
saturation (Fig. \ref{fig:OperatingSystemsSales}) which shows that the
potential market is increasing as the diffusion of this new technology
spreads out. The discussions on the form and the parameter values of
competition together with the dynamic market growth will be the focal
point of this section.

\subsubsection{\label{SectionCatagoryLevel}Category level diffusion of innovation}

\begin{figure}[h]
  \centering
  \begin{subfigure}[b]{0.47\textwidth}
 \includegraphics[width=\textwidth]{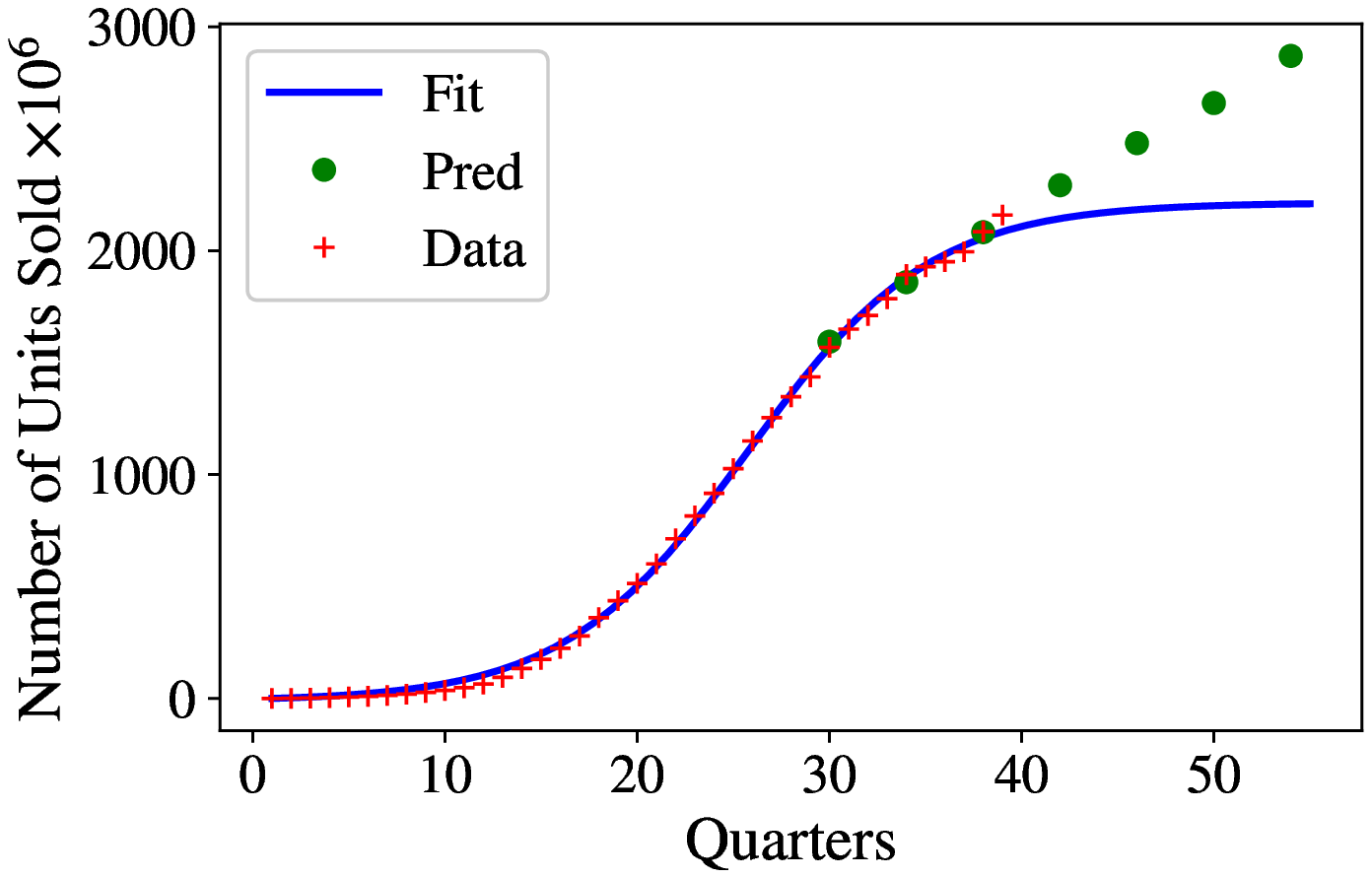}
    \caption{Constant  market size }
        \label{fig:CategoryDynamicMarket_0}
  \end{subfigure}
  \;
  \begin{subfigure}[b]{0.47\textwidth}
    \includegraphics[width=\textwidth]{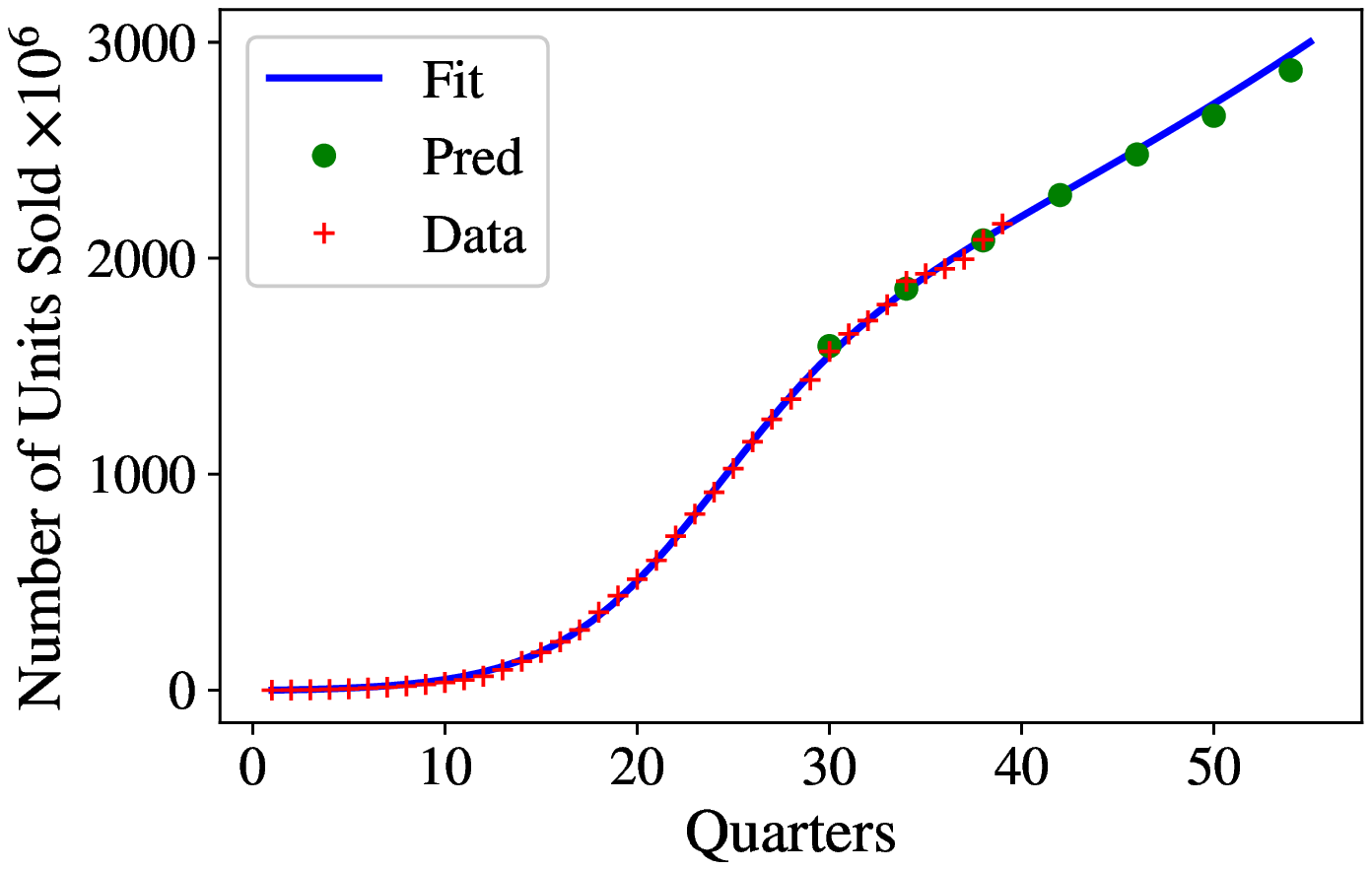}
    \caption{Model I.}
        \label{fig:CategoryDynamicMarket_4}
  \end{subfigure}  
\\
  \begin{subfigure}[b]{0.47\textwidth}
    \includegraphics[width=\textwidth]{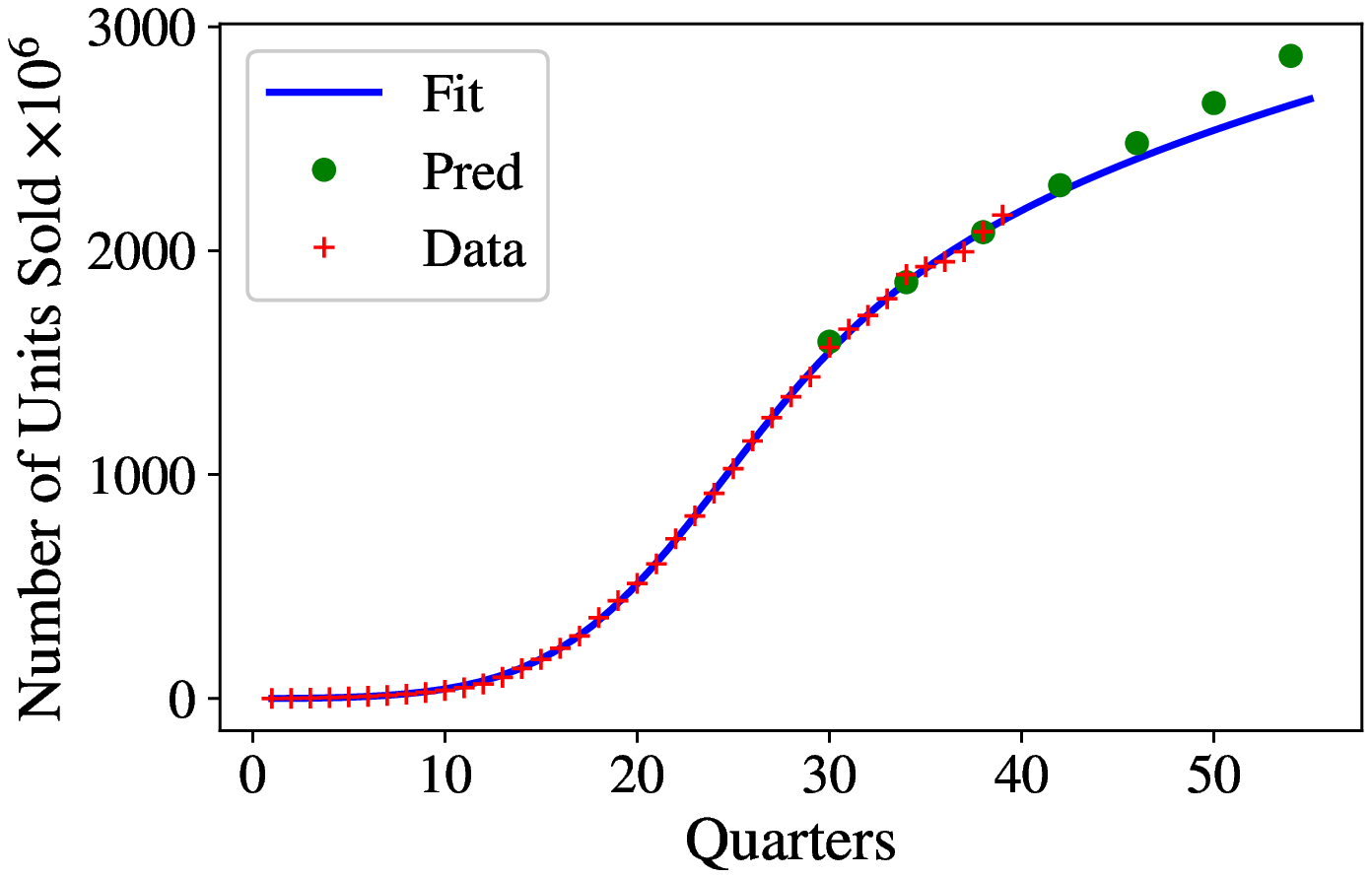}
    \caption{Model II.}
    \label{fig:CategoryDynamicMarket_1}
  \end{subfigure} \;
    \begin{subfigure}[b]{0.47\textwidth}
    \includegraphics[width=\textwidth]{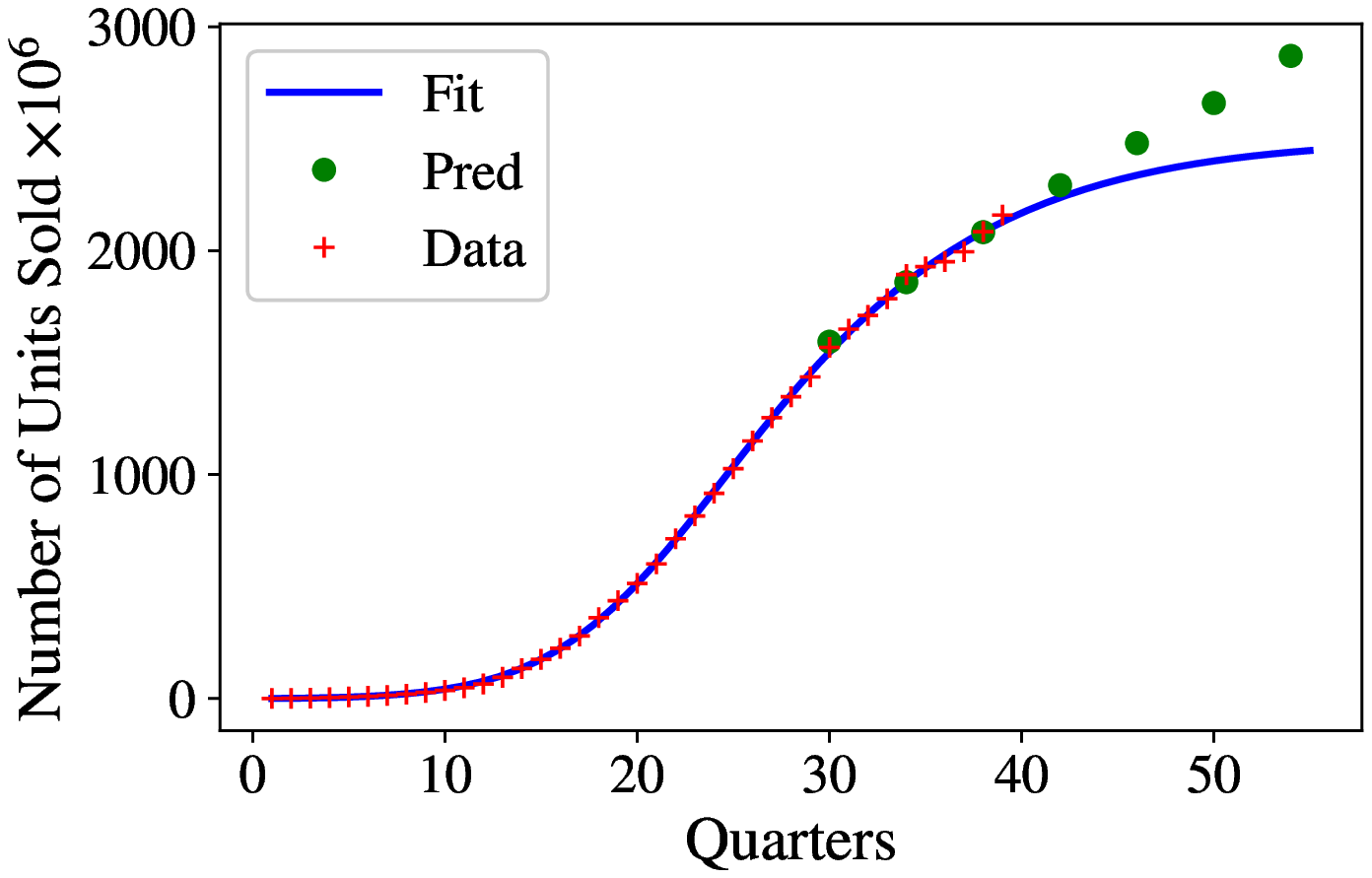}
    \caption{Model III}
    \label{fig:CategoryDynamicMarket_3}
    \end{subfigure}
      \caption{Cumulative smartphone sales modeled by using Bass
    equation with four different  market definitions.}\label{CategoryLevelDiffusionWithDynamicMarket}
\end{figure}

Figure \ref{CategoryLevelDiffusionWithDynamicMarket} show that, the
Bass model fits reasonably well to the existing data. The shortcoming
of the Bass equation with constant market potential (Figure
\ref{fig:CategoryDynamicMarket_0}) is its prediction of saturation in
sales. The saturation is seen immediately after the date where the
data points end. Hence there is a serious discrepancy between the Bass
model predictions and the forecasts on the feature sales potential. To
overcome this shortcoming of the Bass Model three different dynamic
market potential forms are tested (Figures
\ref{fig:CategoryDynamicMarket_4}, \ref{fig:CategoryDynamicMarket_1}
and \ref{fig:CategoryDynamicMarket_3}). The contribution of the
self-reinforcing effect of the dynamic markets~\cite{Guseo:2009} are
also tested. For all dynamic market models, conclusions are the same
for both Bass and Bass with additional correction term (GGM).

$\;$

The resemblance between the existent data and the fits are very good
for all four cases. The fit parameters, standard deviations and
goodness of the fit ($R^2$) values for all four models of market
potential are presented in table~\ref{CategoryData}. In the
table, $M_I$,$ M_{II}$ and $M_{III}$ indicate dynamic market models
defined by the equations, \eqref{MarketGrowth1},\eqref{MarketGrowth2}
and \eqref{MarketGrowth3} respectively. As far as the cumulative sales
data, until the year 2017, are concerned, constant and time-dependent
growing market potential models fits are very satisfactory.  Two main
criterias, standard deviation and $R^2$ comparisons indicate that the
dynamic market potential case is slightly better.

\begin{table} 
  \begin{tabular}{|c|c|c|c|c|}
    \hline
          & $ M = {\rm const}   $             & $ Model\; 1\; (M_I)   $          & $ Model\; 2 \;(M_{II})  $            & $ Model\; 3\;  (M_{III}) $  \\  
\hline
$p$       & $ 1.1543\times 10^{-3} $     & $ 1.3251\times 10^{-3} $        & $ 2.1652\times 10^{-3} $         & $ 2.7556\times 10^{-3} $  \\
          & $ \pm 8.3080\times 10^{-5} $ & $ \pm 8.1876\times 10^{-5} $    & $ \pm 1.2786\times 10^{-4} $     & $ \pm 9.2427\times 10^{-4} $ \\
\hline
$q$       & $ 2.0899\times 10^{-1} $     & $ 2.3685\times 10^{-1} $        & $ 2.1568\times 10^{-1} $  & $ 2.2184\times 10^{-1} $ \\
          & $ \pm 4.5457\times 10^{-3} $ & $ \pm 5.7023\times 10^{-3} $    & $ \pm 6.1231\times 10^{-3} $     & $ \pm 9.6020\times 10^{-3} $ \\
\hline
$M_0$     & $ 2.2144\times 10^{+3} $     & $ 1.0099\times 10^{+3} $        & $ 3.9004\times 10^{+3} $         & $ 2.5013\times 10^{+3} $ \\
          & $  \pm 2.1684\times 10^{+1}$ & $ \pm 7.4212\times 10^{+1} $     & $ \pm 6.2595\times 10^{+2} $    & $ \pm 1.7755\times 10^{+2} $ \\
\hline
$\delta$  & $  -                 $   & $ 1.9810\times 10^{-2} $         & $ 2.1116\times 10^{-2} $  & $ -    $ \\
          & $  -                 $   & $ \pm 1.8909\times 10^{-3} $     & $ \pm 5.3071\times 10^{-3} $ & $ - $ \\
\hline
$p_M$     & $  -             $      & $ -                  $ & $ -   $                & $ 2.6312\times 10^{-3} $ \\
          & $ -              $    & $ -                   $ & $ -   $               & $ \pm 2.4338\times 10^{-3} $ \\
\hline
$q_M$     & $ -                  $   & $  -                 $ & $ -  $               & $ 1.2489\times 10^{-1} $ \\
          & $ -                  $   & $  -                    $ & $  -       $           & $ \pm 4.6963\times 10^{-2} $ \\
\hline
$\sigma$  & $   25.5307 $             & $ 15.1516 $                &    $ 13.74302 $                     & $ 14.1204 $ \\
$R^2$     & $   0.9989 $               & $ 0.9996 $    & $  0.9997 $                          & $ 0.9997 $ \\
\hline
  \end{tabular}
  \caption{Fit parameters of Bass Model with four different dynamic market potential definitions.\label{CategoryData}}
  \end{table}

One can argue that dynamic market potential definitions increase the
number of parameters of the model. Therefore the existence of extra
parameters may increase the flexibility of the fit function and may
result in smaller standard error and better $R^2$ value. For all three
different dynamic market models (Equations
\eqref{MarketGrowth1},\eqref{MarketGrowth2} and \eqref{MarketGrowth3})
F-tests are performed (Table~\ref{CategoryFTests}).  The F-test values
give clear indication that the Bass model with the dynamic market
potential models fit better than the constant market potential.

\begin{table}
  \centering
\begin{tabular}{|l|l|l|l|}
    \hline
    $F_{M_I}$  &  $F_{M_{II}}$ & $F_{M_{III}}$   & $F_{M_{IV}}$ \\
      \hline
      $44.46$   &  $89.24$    & $41.84$      & $67.21$    \\
        \hline
  \end{tabular}
  \caption{F-test results : Model comparisons between the constant and
    dynamic market potential solutions of the Bass model.\label{CategoryFTests}}
  \end{table}

From the marketing point of view, steep increase in the quarterly
sales is an indication of growing market size. Moreover, global
smartphone users data and predictions on the future number of users
indicate that the cumulative sales are far from a saturation
situation~\cite{WorldWideSmartphoneUsers}.

$\;$

The main test of the dynamic market models is their prediction power.
All three dynamic models are put to test for predicting future
sales. The fits to the existing quarterly sales data extrapolated
until the year 2020. The extrapolations are compared with the existing
prediction data.  Figure \ref{CategoryLevelDiffusionWithDynamicMarket}
show the fit and extrapolation regions of all market potentials. All
of the dynamic market models exhibit improvements over the constant
market case. Never the less, first two models, (Model I, figure
\ref{fig:CategoryDynamicMarket_4} and Model II, figure
\ref{fig:CategoryDynamicMarket_1}) are better for predicting near
future sales. As a conclusion of this subsection it can be said that
the best agreement between the predictions and the solution of the
Bass equation can be reached when the potential market exhibits
exponential growth with time ($M_I$).

\subsubsection{Product level diffusion of innovation}

To compare the effects of different dynamic market potentials under
competition, four different market potential are presented in the
context of three diffusions of innovation models.  In this section
three models, which are introduced by
Kirshnan-Bass-Kumar~\cite{Krishnan:2000}(KBKM),
Guseo-Mortarino~\cite{Guseo:2015} (GMM) and
Libai-Muller-Peres~\cite{Libai:2009} (LMPM), are compared. The model
parameters together with the parameters of dynamic market potentials
require an extra care for the fitting procedure. Since the dimensions
of the parameter space is high, to avoid the local minima, a random
search is performed to choose the lowest standard deviation.  The
lowest standard deviation value is chosen from 500 minimization runs
with random initial parameter sets are used to determine the lowest
standard deviation value. The parameter set which minimizes the
standard deviation is presented as the parameter set in the
results. The number of available data points are sufficient for the
estimation of the parameters. Newer the less, sophisticated
statistical analysis require more data points for the given number of
parameters. Despite all the efforts, the error estimation using
convolution matrix or bootstrap methods can not give satisfactory
results. Hence parameter error estimates are not presented in this
subsection.

$\;$

Solutions of coupled differential equations, for all three models are
obtained by using SciPy odeint packet if no analytical form
exist~\cite{Guseo:2015}.  The models of diffusion of innovation
(KBKM,GMM,LMPM) are very successful fitting the existing quarterly
cumulative sales data. The discrepancy starts with the prediction of
the future sales. To discuss the effects of the dynamic
market potentials, all three diffusion models are solved by using
alternative market potential forms. Results are very similar,
particularly for the graphical representation of the data and fits are
almost indistinguishable up to the end of quarterly sales data which
is the first quarter of 2017.

\begin{figure}[h!]
  \centering
  \begin{subfigure}[b]{0.47\textwidth}
    \includegraphics[width=\textwidth]{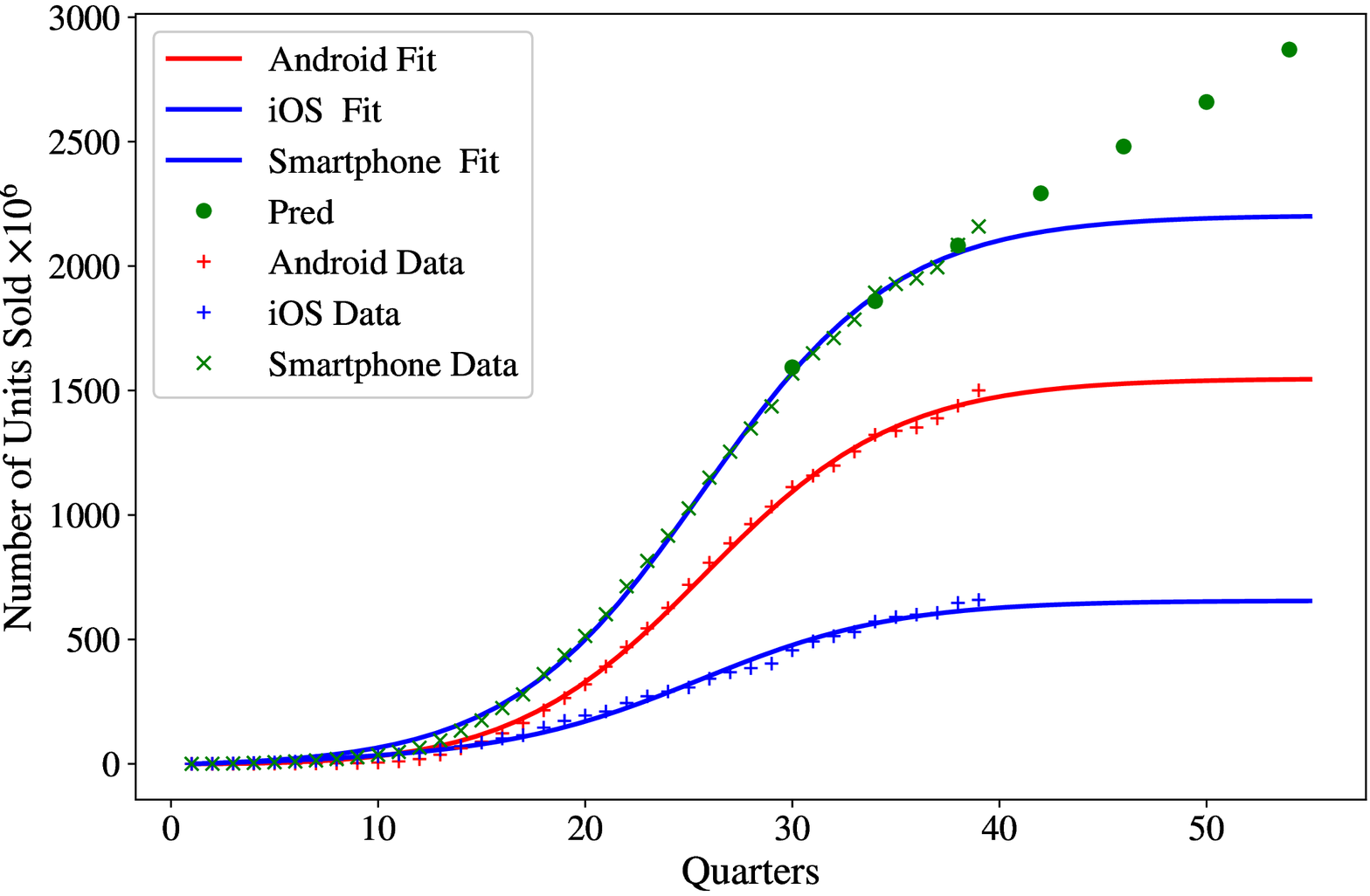}
    \caption{Constant size market case.}
    \label{fig:CategoryProducModel0}
  \end{subfigure} \;
  \begin{subfigure}[b]{0.47\textwidth}
    \includegraphics[width=\textwidth]{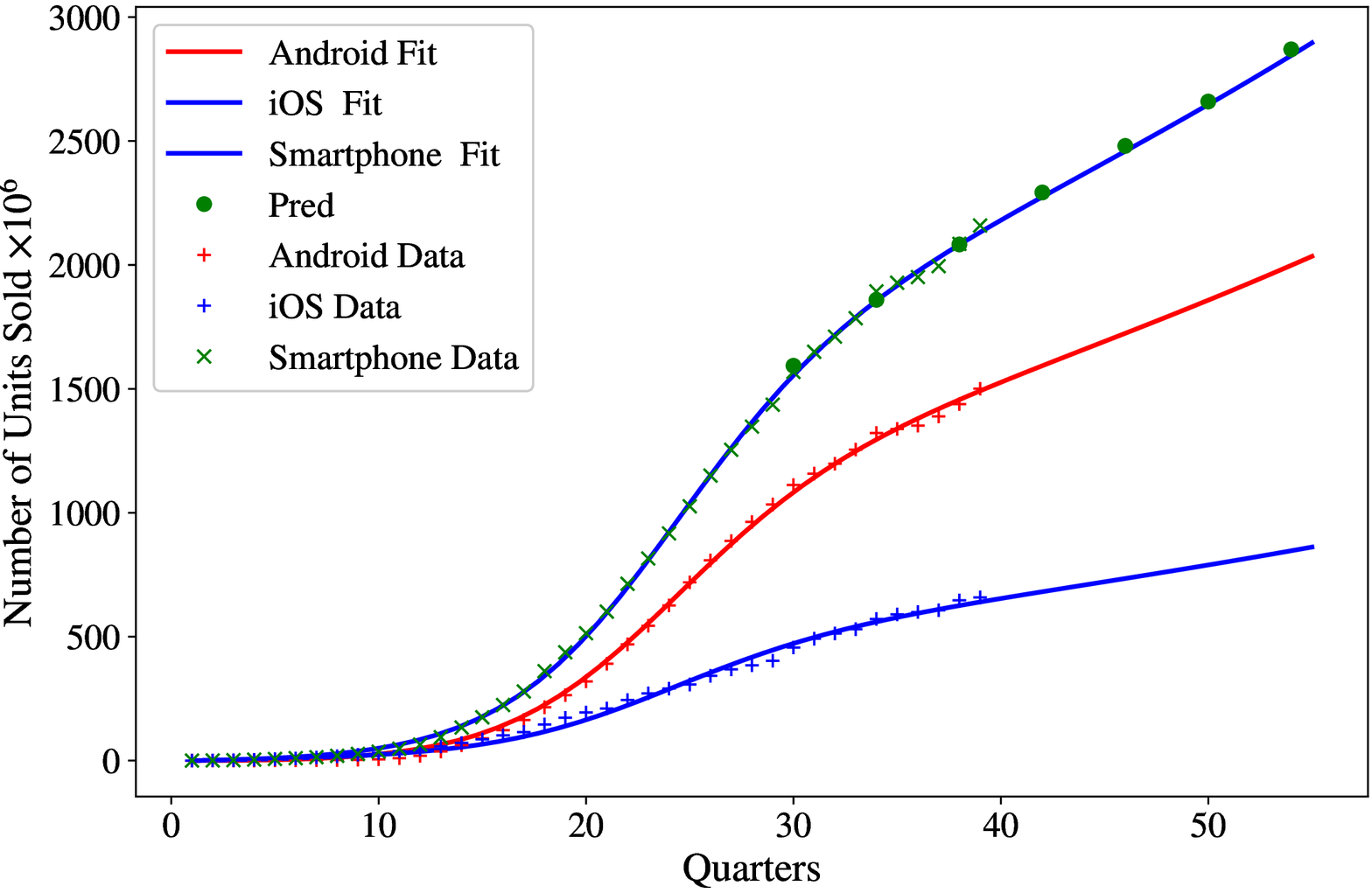}
    \caption{Dynamic market model 1.}
    \label{fig:CategoryProducModel4}
  \end{subfigure}  
\\
  \begin{subfigure}[b]{0.47\textwidth}
    \includegraphics[width=\textwidth]{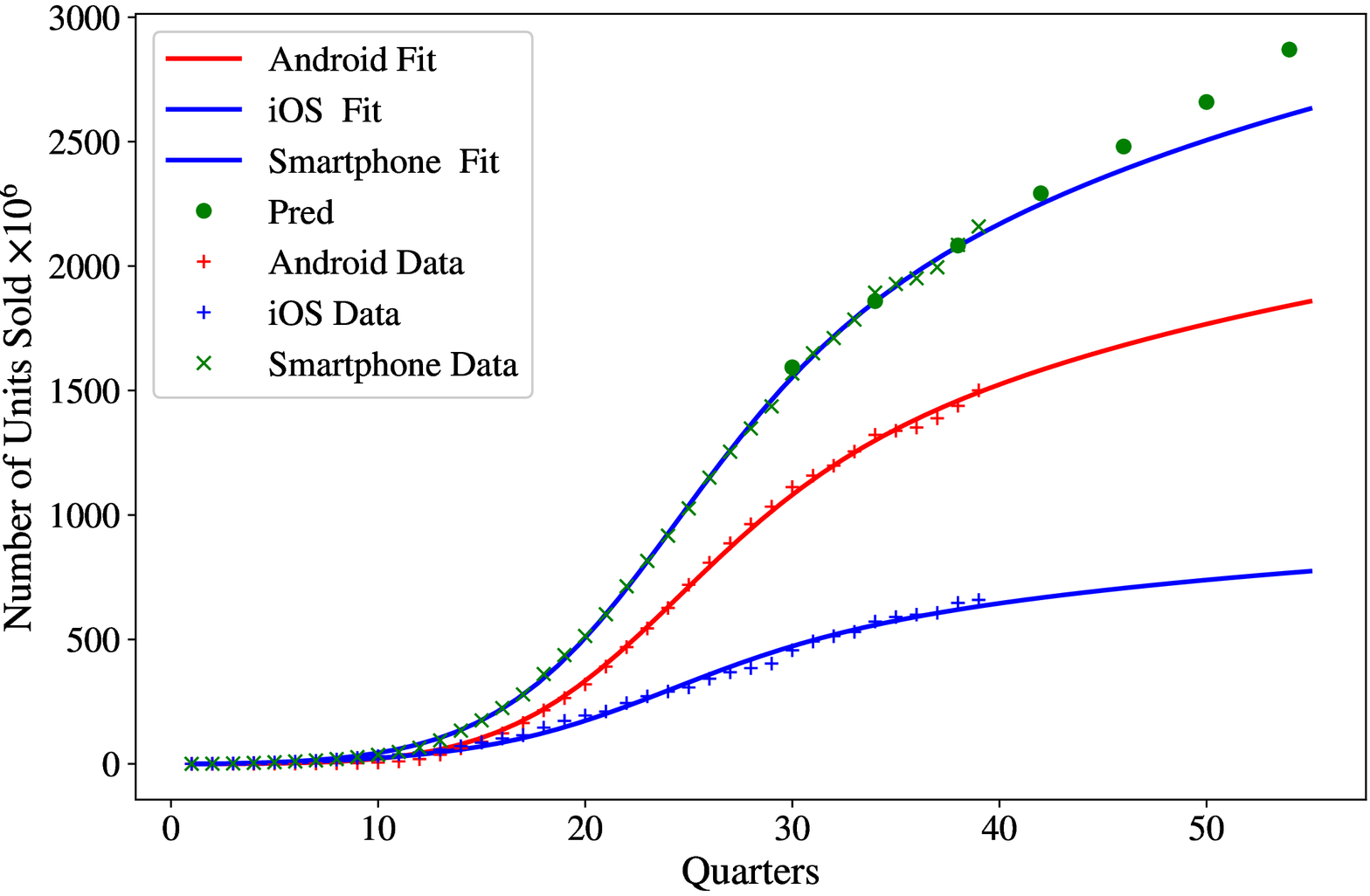}
    \caption{Dynamic market model 2.}
    \label{fig:CategoryProducModel1}
  \end{subfigure}\;  
  \begin{subfigure}[b]{0.47\textwidth}
    \includegraphics[width=\textwidth]{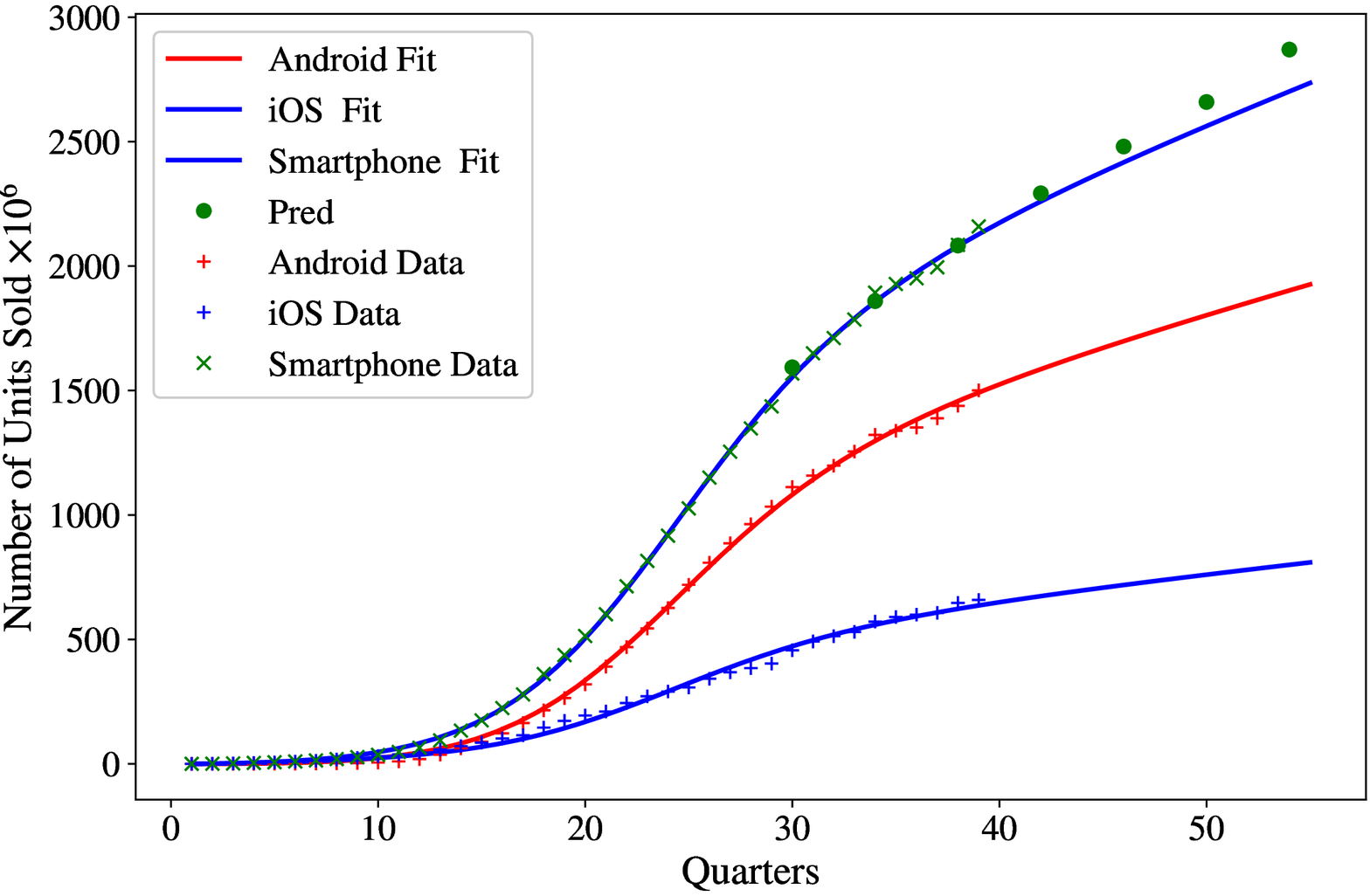}
    \caption{Dynamic market model 3.}
    \label{fig:CategoryProducModel3}
  \end{subfigure} 
  \caption{Cumulative smartphone sales, modeled by using
    Krishnan-Bass-Kumar model~\cite{Krishnan:2000}, with the
    assumptions, a) the potential market size is constant, b,c,d)
    dynamic market potential cases, $M_I,M_{II}$ and
    $M_{III}$. }\label{CategoryProductDiffusion1}
\end{figure}

\begin{figure}[h!]
  \centering
  \begin{subfigure}[b]{0.47\textwidth}
    \includegraphics[width=\textwidth]{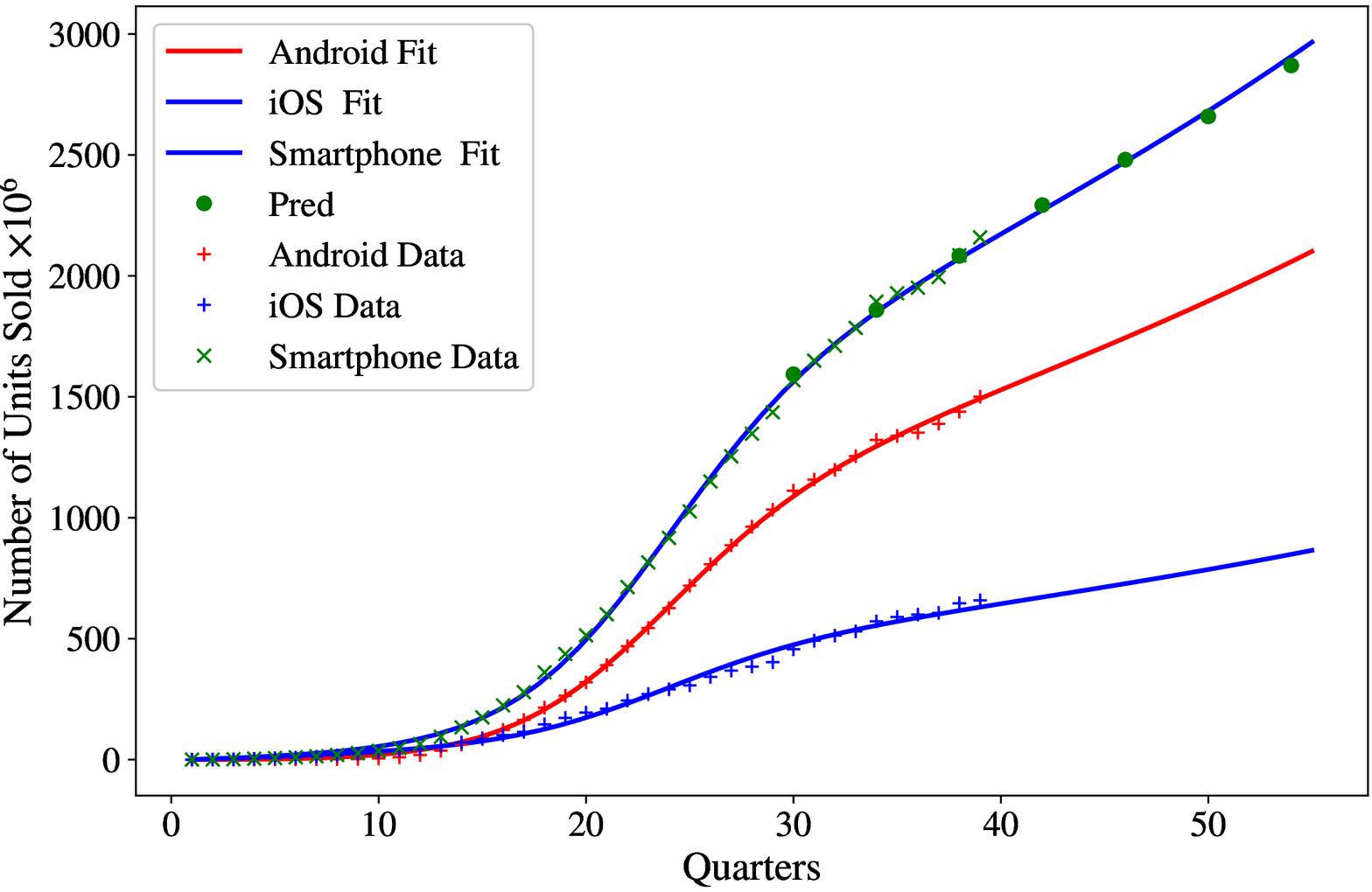}
    \caption{L Model with six parameters.}
    \label{fig:CategoryProducModel2}
  \end{subfigure}  
  \begin{subfigure}[b]{0.47\textwidth}
    \includegraphics[width=\textwidth]{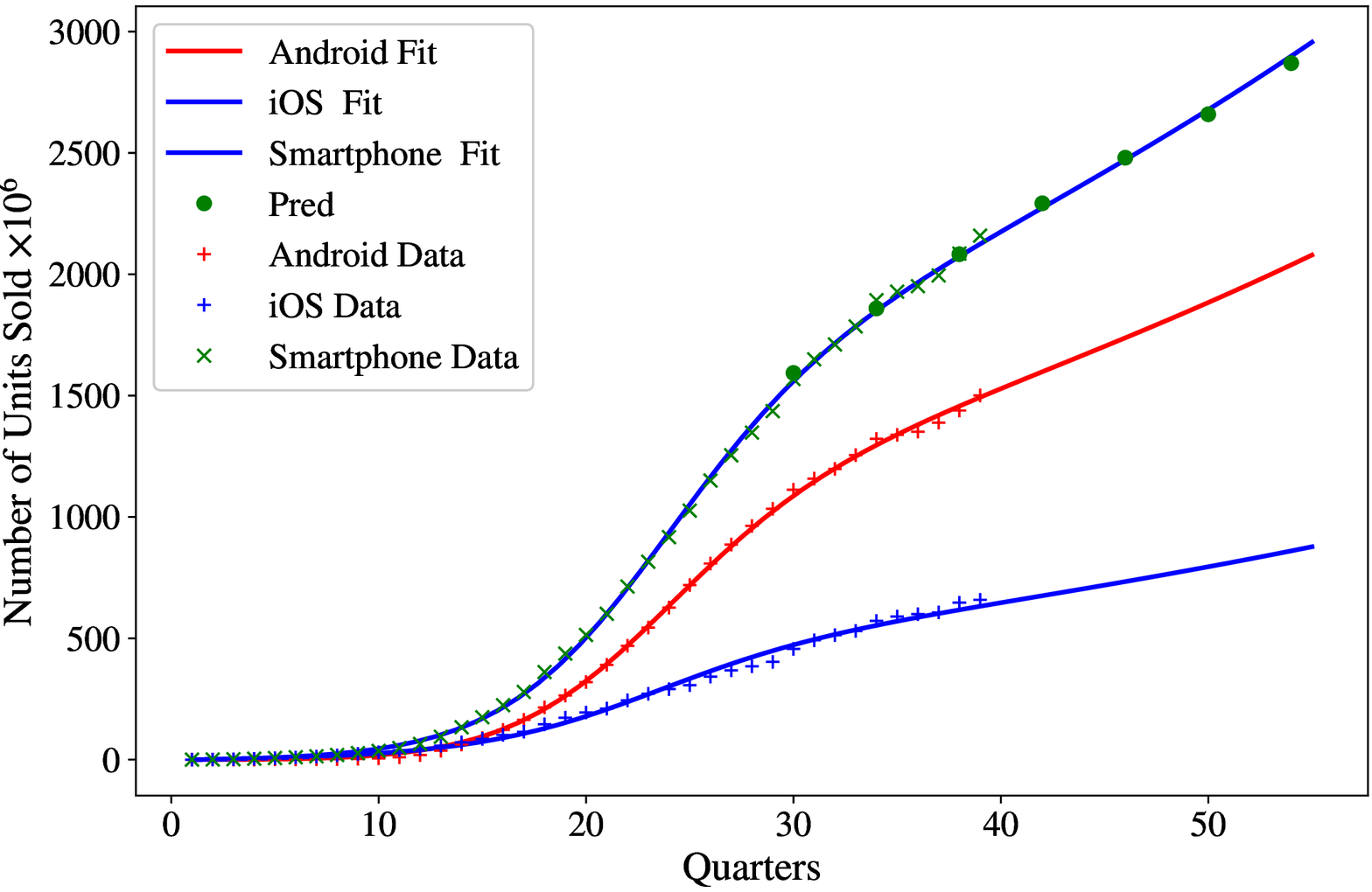}
    \caption{GMM  with 5 parameter.}
    \label{fig:CategoryProducModel3}
  \end{subfigure} 
  \caption{Predictions of cumulative smart phone sales (under
    exponentially growing market assumption) a) LMPM, b) GMM. }\label{CategoryProductDiffusion2}
\end{figure}

As an illustrative example, the results of Kirshnan-Bass-Kumar model
will be presented in detail, later the other two models, GMM and LMPM,
will be added to discussions when additional arguments contribute
further.  Figure \ref{CategoryProductDiffusion1} shows the solutions
of the KBKM with constant and dynamic market potential definitions.
Figure~\ref{fig:CategoryProducModel0} shows the characteristic S-shape
Of the diffusion of the innovation curve. With constant market
potential approximation, the maximum size of the market is determined
as $2.2077\times 10^{+9}$ costumers. This is approximately one-third
of the world population and this number is the estimated/observed
number of smartphohe users in
2017~\cite{AndroidAndiOS,NumberOfActiveiOSUsers,NumberofActiveAndoidUsers}. Despite
such a wide usage, still smartphone market is growing globally. Hence,
market potential forecasts does not agree with obtained saturation
level.  Second subfigure, (Figure \ref{fig:CategoryProducModel4})
shows the results obtained by using exponentially growing market
potential. This model is unrealistic in the long run since no market
can grow indefinitely. Never the less, it is in very good agreement
with the existing data and the short term predictions on the number of
future users.  The fit to the cumulative quarterly sales data agrees
well, $\chi^2= 37.6642 $ and $R^2=0.9979$, moreover when the fit is
extended until the year 2020, the agreement with the predictions are
remarkably good ($\chi^2=12.2141$ and $R^2=0.9992$). The other two
dynamic Market potential models are also exhibit agreement with the
existing quarterly sales data (Figures \ref{fig:CategoryProducModel2}
and \ref{fig:CategoryProducModel3}) but both predict saturation
immediately after 2017. Hence, only exponentially growing market
approximation give correct information for the short-term sales
prediction.

$\;$

Table \ref{BassKumarTable} show parameter values, standard deviations,
r-squared values of the fits for all four different market potential
forms. The KBKM fits indicate that the leading effects on the category
level sales are the internal influences. As a consistency check, the
values obtained for the category level parameters, $p_c$ and $q_c$ (in
equation \eqref{ExtendedBass}) are also in accord with the parameter
values obtained from the category level sales
studies (Table ~\ref{CategoryData}). The product level sales are governed
by two sets of parameters which add up to the category level parameter
values. Table \ref{BassKumarTable} indicate that for KBKM, the
external influences play no role in the sales of the first product
(Android OS). On the other hand a considerable number of consumers buy
the second product (iOS) as soon as it is announced. The effects of the
internal influences are larger for the first product (Android) than the
second product, (For constant Market size case, $p_1 \sim
p_c$). Effects of the internal influences are larger for the fist
product (Android) than the second product, which indicate that hands
on experience and personal interactions play crucial role in the
spread of first product. Even for the dynamic market case this
conclusion does not change.

\begin{table}
\centering
\scalebox{0.9}{
  \begin{tabular}{|r|r|r|r|r|}
    \hline
Parameter  & $  M0  $ & $     M4  $ & $  M1  $ & $ MG $ \\
  \hline
$p_c  $ & $1.1308\times 10^{-3}  $ & $ 1.1715\times 10^{-3}  $ & $ 2.0493\times 10^{-3}  $ & $  2.2960\times 10^{-3}  $ \\
  \hline
  $q_c  $ & $2.1047\times 10^{-1}  $ & $ 2.5453\times 10^{-1}  $ & $ 2.1569\times 10^{-1}  $ & $ 2.1766\times 10^{-1}  $ \\
  \hline
  $p_1  $ & $0.0000\times 10^{+00}  $ & $ 0.0000\times 10^{+00}  $ & $ 0.0000\times 10^{+00}  $ & $ 0.0000\times 10^{+00}  $ \\
  \hline
  $q_1  $ & $1.5213\times 10^{-1}  $ & $ 1.8161\times 10^{-1}  $ & $ 1.5825\times 10^{-1}  $ & $ 1.6025\times 10^{-1}  $ \\ 
  \hline
  $M_0  $ & $2.2076\times 10^{+3}  $ & $ 1.1562\times 10^{+3}  $ & $ 3.4796\times 10^{+3}  $ & $ 2.5412\times 10^{+3}  $ \\
  \hline
  $\delta_M  $ & $-           $ & $ 1.8163\times 10^{-2}  $ & $ 2.5320\times 10^{-2}  $ & $ -  $ \\
  \hline
  $p_M   $ & $  -   $ & $ -              $ & $ -            $ & $ 4.8378\times 10^{-3} $ \\
  \hline
  $q_M   $ & $  -  $ & $ -   $ & $ -     $ & $9.9715\times 10^{-2}  $ \\
\hline
$\sigma$  & $  41.4162 $ &    $ 37.6642 $  & $ 36.6009 $   & $ 37.6643 $ \\
$R^2$     & $  0.9974$    & $ 0.9979 $ & $ 0.9980 $        & $ 0.9979 $ \\
%$R^2_{Reduced}$ & $ 0.9978 $  & $ 0.9983 $  & $ 0.9984 $    & $ 0.9984$ \\
\hline
  \end{tabular}
}
  \caption{Bass-Kumar model parameters for four different Market potantial forms\label{BassKumarTable}}
  \end{table}

Competition between the rival products is not openly pronounced in the
Kirshnan-Bass-Kumar model. The other two models have some extra
parameters designated to measure the contribution of the competition.
The LMPM model contains two external, $p_1$ and $p_2$ and four
internal parameters, $q_1,q_2,c_1$ and $c_2$. The internal parameters,
$c_1$ and $c_1$ are related with the cross-brand influences.  GMM has
two external parameters, $p_1$ and $p_2$, which have exactly the same as KBKM and LMPM,
three parameters, $q_1,\; q_2$ and $\delta$, related with the internal effects which have slightly
different meaning. The parameter $\delta$
is employed to represent the cross-brand effects.  Comparing internal parameters
of LMPM and GMM, the correspondence can be given as, $ c_1^{LMPM} =
q_1^{GMM}$, $ c_2^{LMPM} = q_2^{GMM} - \delta$. Tables, \ref{LTable}
and \ref{GuseoTable} show the parameter values obtained using dynamic
market growth with exponential time dependence $M_I$. Both models
predict positive cross-brand contribution. Both operating systems
benefit from the existence of the other operating system.

\begin{table}
\scalebox{0.9}{
\begin{tabular}{|l|l|l|l|l|l|}
\hline
$p_1$   &$q_1$&$p_2$&$q_2$&$c_1$&$c_2$ \\
  \hline
 $ 0.0000 $& $2.6573\times 10^{-1}$&  $2.8145\times 10^{-3}$  &  $1.4201\times 10^{-4}$ &  $5.9847\times 10^{-2}$ &   $1.0758\times 10^{-1}$\\
\hline
\end{tabular}
}
\caption{ LMPM parameters\label{LTable}}
\end{table}

\begin{table}
\centering
\scalebox{0.9}{
\begin{tabular}{|l|l|l|l|l|}
\hline
$p_1$    &    $q_1$   &  $p_2$ &  $q_2$ & $\delta$ \\
  \hline
$0.0000$ & $ 1.0721\times 10^{-1}  $   &   $ 1.1247\times 10^{-3}  $  & $ 1.4072\times 10^{-1} $ & $ 1.1136\times 10^{-1}$ \\
\hline
\end{tabular}
}
\caption{GMM parameters\label{GuseoTable}}
\end{table}

For both models, the fitted parameters of the dynamic market potantial
is given in table \ref{MassTable}.

\begin{table}
\centering
\scalebox{0.9}{
\begin{tabular}{|l|l|l|}
\hline
        &  $M_0$   &            $\delta_M$\\
  \hline
  LMPM  & $1.0607\times 10^{3}$ &  $2.0029\times 10^{-2}$\\
    \hline
GMM   & $1.0076\times 10^{3}$  & $1.9565\times 10^{-2}$ \\
 \hline
\end{tabular}
}
\caption{Exponentially growing  dynamic market potential parameters for LMPM and GMM.\label{MassTable}} 
\end{table}

Figure \ref{CategoryProductDiffusion2}, shows category level diffusion
modelled by using LMPM and GMM models. Both models accompany
exponentially growing dynamic market potential. The chi-square and r-square values give the quality of the fit for LMPM, $\chi^2 = 39.2933,\;\; R^2_1=0.9979$, and for GMM $\chi^2=37.0053,\;\; R^2_1=0.9980$.

% c1 = q1 = 1.0e-1
% c2 = q2 - d = 1,4 e-1 - 1.1 e-1 = 0.3 e-1
%q1 = (q1 + d)  = 1.0 e-1 + 1.1 e-1 = 2.1 e-1
%q2 = q2        = 1.4 e -1

Figure \ref{CategoryProductDiffusion2} also show that for both models
exponential growing market model succesfully predict the values
obtained in ref ~\cite{WorldWideSmartphoneUsers}.  The extrapolation of the
fit matches the prediction data very well, for LMPM. $\chi_2 =
22.3559, R^2_2=0.9975$ and for GMM, $\chi_2 = 17.6432, R^2_2=0.9983$.

%Function = Guseo_1, Market = Mass_4 
%$\chi_1 = 37.005348, R^2_1=0.998063, R^2_1_{Adj}=0.998420$
%$\chi_2 = 17.643222, R^2_2=0.998365, R^2_2_{Adj}=0.998842$
%Analysis >python Statistics.py Coupled_Bass3 Mass_4
%Function = Coupled_Bass3, Market = Mass_4 
%$\chi_1 = 39.293348, R^2_1=0.997885, R^2_1_{Adj}=0.998330$
%$\chi_2 = 22.355985, R^2_2=0.997521, R^2_2_{Adj}=0.998348$

\section{Conclusions}

The Bass model has a  great success in explaining diffusion
mechanisms of an extensive range of products. Despite its success, the Bass model cannot accommodate
economic and social parameters which have a direct influence on the
sales and market success of a new product. Attempts to extend the Bass
model to consider social and economic parameters
complicates the model. Since the parameters such as price, utility, social
prestige, competition-driven popularity have different weights and
importance for different product categories. Extra complications restrict the prediction power of the model. The most notable effect
of all these parameters is to increase or decrease the demand. Hence,
the above mentioned economic and social parameters are the driving
forces of the dynamics of market growth. 

$\;$

In this work answers to three main conceptual questions are searched:
a) what are the relative importance of the external and internal
influences of a high technological innovation?, b) How one can predict the sales for the near future? c) What is the effect of the
competition in smartphone diffusion of innovation? Diffusion of
innovation model with constant market size may give reasonable answers to these questions if the model is applied to the existing data after the product diffusion is reached its saturation level. The main
concern, if one can predict future sales by using Bass or extended
Bass model must accommodate the information on the dynamics of market growth.

$\;$

In this work, the diffusion of smartphones operating systems is
studied by using models presented in section~\ref{Models} and the
existing data.  Diffusion of smartphones has an exceptional place in
the diffusion of innovation studies~\cite{DeGusta:2012}. First of all,
there is no other technological innovation which substitutes so
many different technological appliances. Secondly, the social impact of smartphone technology has no match. Thirdly, smartphone technology is a living technology; frequent introduction of new versions with
advanced features wet the appetite of the potential adopters and
boost the sales. Two operating systems, namely Android and iOS dominate the smartphone markets.  Hence smartphone operating systems
provide a laboratory for competitive market studies. 

$\;$

The relative importance of the external and internal influences are studied by fitting the quarterly sales data with the solutions of
different diffusion of innovation models. It is observed that both in the category and the product level, external influences are quite
insignificant.  Instead, internal influences are the leading effects
in both category and the brand level sales. This observation is related to the product itself. Smartphones create a new type of
social interaction, namely social media, which enabled individuals to
share experiences and opinions. The social media created a different
type trusted information source which is the most influential factor
in decision-making processes.

$\;$

The near future sales predictions can be made by introducing a dynamic
market potential. Any market potential with saturation can be used if
the limits of the market size are known. On the other hand, for short
term predictions market models with time-dependent growth fits better
Figure \ref{CategoryProductDiffusion1}.

$\;$ 

High competition helps the growth of the market by increasing the awareness of the potential adopters. It is seen from the inter-product
competition terms (Tables \ref{LTable} and \ref{GuseoTable}) that
positive the existence of competitors contribute positively to the
sales of both brands.

\end{document}